# CALCULATION OF PHASES OF *NP*- SCATTERING UP TO $T_{lab}$=3 GeV FOR Reid68 AND Reid93 POTENTIALS ON THE PHASE-FUNCTION METHOD

*V. I. Zhaba*

*Uzhhorod National University, 88000, Uzhhorod, Voloshin Str., 54*

For calculation of the single-channel nucleon-nucleon scattering a phase-functions method has been proposed. Using a phase-functions method the following phase shifts of *np*- scattering numerically for $^1S_0$-, $^1P_1$-, $^3P_0$-, $^3P_1$-, $^1D_2$-, $^3D_2$-, $^1F_3$-, $^3F_3$-, $^1G_4$-, $^3G_4$- states are calculated. The calculations has been performed using realistic nucleon-nucleon Reid68 and Reid93 potentials. Obtained phase shifts for energy up to 350 MeV are in good agreement with the results obtained in the framework of other methods. Using the obtained phase shifts we have calculated the full cross-section *np*- scattering.

## 1. INTRODUCTION

Based on the experimentally observed values of the scattering cross-section and energies of transitions we get information about the scattering phases and amplitudes in the first place, than about the wave functions, which are the main object of the research in a standard approach. In other words, not the very wave functions are being observed in the experiment, but their changes caused by the interaction [1,2]. It is therefore of interest to obtain and use the equations directly connecting the phases and scattering amplitudes with the potential without finding the wave functions.

The precise solution of the scattering problem aiming at calculation of the scattering phase is possible only for individual phenomenological potentials. When realistic potentials are used, the phases of scattering are roughly calculated. This is due to the use of physical approximations or numerical calculation.

In the last 10 years has increased the interest in nucleon-nucleon scattering in the framework of chiral perturbation theory [3,4], a coherent theoretical-field approach [5], for partial wave analysis below the threshold for formation of the pion [6]. Also phase of scattering get through supersymmetry and factorization [7], through N/D method for calculation of partial waves for elastic *NN*- scattering [8] or renormalisation *NN*- interaction potential for the chiral twopion exchange [9].

The methods of solving the Schrödinger equation with the aim of obtaining the scattering phases include: the method of successive approximations, the Born approximation, the phase-functions method (PFM), and others. PFM appeared convenient for solving many tasks of atomic and nuclear physics.

When applied to problems of nucleon-nucleon scattering the main and foremost advantage of PFM is such that PFM allows to obtain the scattering phase, without finding the wave functions as solutions of the Schrödinger equation. Due to a phase equation there is a direct relationship between the scattering phase shift and the interaction potential.

This paper deals with the calculation of the phase shifts of *np*- scattering in the relevant spin configurations for the realistic phenomenological nucleon-nucleon potentials Reid68 [10] and Reid93 [11] by using PFM.

## 2. THE PHASE-FUNCTIONS METHOD

PFM is a special method to solve the radial Schrödinger equation

$$u''_l(r) + \left(k^2 - \frac{l(l+1)}{r^2} - U(r)\right) u_l(r) = 0, \quad (1)$$

which is a second order linear differential equation. In the formula (1) the value $U(r) = \frac{2m}{\hbar^2}V(r)$ - is the renormalized interaction potential, *m* - the reduced mass. PFM it is quite convenient for obtaining scattering phases, because this method does not require calculating radial wave functions of scattering problem in a wide range firstly and then finding these phases by their asymptotics.

The standard method of calculating the scattering phases is a solution of the Schrödinger equation (1) with the asymptotic boundary condition. PFM is the transition from Schrödinger equation to the equation for the phase function. For that purpose one should change [1,2]:

$$u_l(r) = A_l(r)\left[\cos\delta_l(r) \cdot j_l(kr) - \sin\delta_l(r) \cdot n_l(kr)\right]. \quad (2)$$

The two new introduced functions $\delta_l(r)$ and $A_l(r)$ are the corresponding scattering phases and normalization constants (amplitudes) of wave functions for scattering on a determined sequence of truncated potentials. $\delta_l(r)$ and $A_l(r)$ are called a phase and an amplitude function according to their physical content. The term "phase function" was first used in the paper by Morse and Allis [12]. Equation for phase function with the initial conditions are:

$$\delta'_l(r) = -\frac{1}{k}U(r)\left[\cos\delta_l(r) \cdot j_l(kr) - \sin\delta_l(r) \cdot n_l(kr)\right]^2, \quad (3)$$

$$\delta_l(0) = 0.$$

The phase equation was obtained for the first time by Drukarev, and then independently in the works of Bergmann, Calogero and Zemach. A special case of the phase equation (3) at $l=0$ has been used by Morse and Allis at examination of problem of $S$- scattering of slow electrons on atoms [12].

## 3. CALCULATIONS OF PHASE SHIFTS AND DISCUSSION OF RESULTS

By the phase-functions method it has been numerically obtained the phase shifts of $np$- scattering for $^1S_0$-, $^1P_1$-, $^3P_0$-, $^3P_1$-, $^1D_2$-, $^3D_2$-, $^1F_3$-, $^3F_3$-, $^1G_4$-, $^3G_4$-states. The masses of nucleons have been chosen as: $M_p$=938.27231 MeV; $M_n$=939.56563 MeV. The Runge-Kutta method of the fourth order [13] was chosen as the numerical method of solving the phase equation (3). Program code for numerical calculations was written in the programming language FORTRAN. The phase shifts were obtained with a precision of 0.01 for optimized selection steps for numerical calculations. The phase shifts were at an output of the phase function $\delta_l(r)$ on an asymptotics at $r>25$ fm. The values of phase shifts are shown in Fig. 1-3. The phase shifts are specified in degrees. The numerical calculations have been carried out for Reid68 and Reid93 potentials. The interval of energies was 1-3000 MeV.

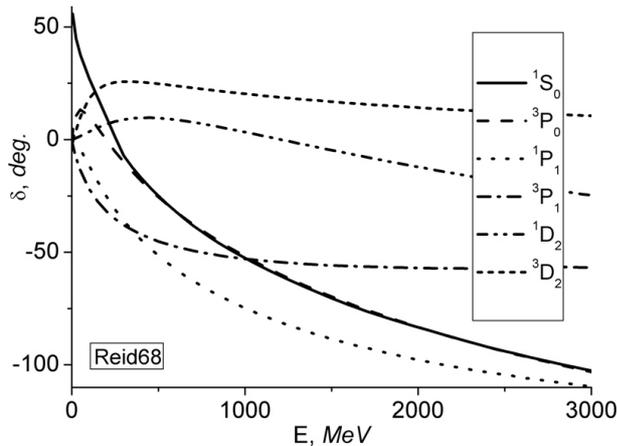

*Fig. 1.* Phase shifts of $np$- scattering for Reid68 potential

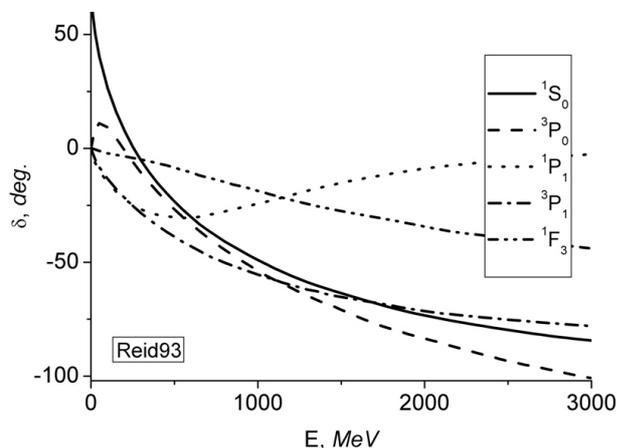

*Fig. 2.* Phase shifts of $np$- scattering for Reid93 potential

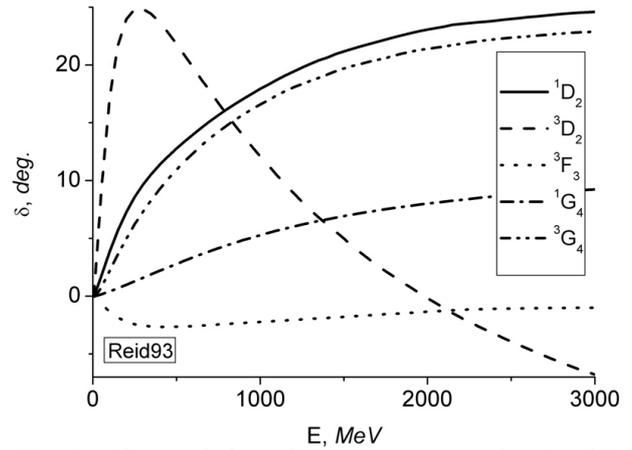

*Fig. 3.* Phase shifts of $np$- scattering for Reid93 potential

For the energy range 1-350 MeV there is good agreement between the phase shifts, obtained on the basis of PFM (outcomes of the given paper) and data's in other papers [10,11]). The discrepancy between the outcomes makes no more than two percent.

It should be noted that in the last 15 years has increased the interest for finding phase shifts at high energies. Unfortunately, in the literature considerably less than the existing calculated phase shifts at high energies for the Reid potentials. But for other potential models are the phase shifts in a wide energy region. So the phase shifts calculated up to 1000 MeV for the nucleon-nucleon models Av18, CD-Bonn and N3LO [14], as well as for Arndt, OBEP, Bonn, Nijm-3 and Paris [15]. In [16] shows the results of calculations of the phase shifts up to 1.6 GeV for the potentials of the inversion based on SM94, OSBEP, Av18 and Bonn-B. In addition, according to [17] potentials Nijm-1, Nijm-2, Av18 and the quantum inversion Gelfand-Levitan-Marchenko were expanded as $NN$ optical models. And took into account the analysis of phase shifts by Arndt et al. (SP00, FA00, WI00) from 300 MeV to 3 GeV. In [18] the available phase shifts only up to 1000 MeV for $^1F_3$- state for Nijm-1, Nijm-2, Reid93, Av18 and Bonn potentials, which were extrapolated for high energies.

Besides the results mentioned in the papers, the available phase shifts obtained up to 3 GeV for the relativistic optical model based on the Moscow potential [19] and MYQ2, MYQ3, MY2, SP07 and Graz II potentials [20], and up to 1.2-3.0 GeV for Dirac potential [21] and up to 1.2 GeV for the inversion potential and the Paris potential [22].

The difference between the obtained phase shifts $^1F_3$- state by PFM and to the data in [18] for the Reid93 potential is not more than 5 percent. Eventually the calculated phase shifts for a particular spin configuration at high energies (more 350 MeV) for Reid68 and Reid93 potentials differ in most cases.

If we compare obtained the phase shifts at high energies for Reid potentials by PFM with the data for the other potential models, then obvious difference between them. Of course, this is due to the particular structure of the nucleon-nucleon potentials.

Despite the obvious improvement in the description of the data at energies below 400 MeV, the theoretical calculations so far show some significant systematic shortcomings [22]. In particular, it remains unclear the divergence of the spin observed at momentum transfer below 1 fm$^{-1}$. The derivation of such discrepancies with the data could be attributed to the phenomenological limitations of «bare» nucleon-nucleon potential, especially at higher energies, as well as simplifications in the model for the *NN* effective interaction or to the fact that the optical model potential has been developed only to the lowest order. Above 350-400 MeV differences between the potentials of *NN* interaction become more obvious. At energies more than 400 MeV the *NN* potentials have applications. Therefore, the evaluation of the theory requires a more precise description of the «bare» interaction between two nucleons.

Along with the phase shifts in the problems of scattering one should deal with the scattering amplitudes, *S*- matrix elements and a number of other parameters. Based on the known phases of scattering one can obtain the complete amplitude, the full cross-section and the partial scattering amplitude accordingly [1]

$$F(\theta) = \frac{1}{k}\sum_{l=0}^{\infty}(2l+1)e^{i\delta_l}\sin\delta_l P_l(\cos\theta), \quad (4)$$

$$\sigma = \frac{4\pi}{k^2}\sum_{l=0}^{\infty}(2l+1)\sin^2\delta_l, \quad (5)$$

$$f_l = \frac{1}{k}e^{i\delta_l}\sin\delta_l, \quad (6)$$

where $P_l(\cos\theta)$ - Legendre polynomials, $\theta$ - polar angle.

In paper [23] specified the full cross-section scattering, calculated with the phase shifts at energies 1-350 MeV for potentials Nijmegen group (including for Reid93) and Av18 potential. Using the phase shifts at energies up to 350 MeV, the calculated results for the Wolfenstein parameters as a function of angle C.m. and $E_{LAB}$ are given in [24] for PWA and Av18, as in [25] for Reid93 and NijmII.

The calculation results of the full cross-section of *np*- scattering (5) are presented in Fig. 4 and 5.

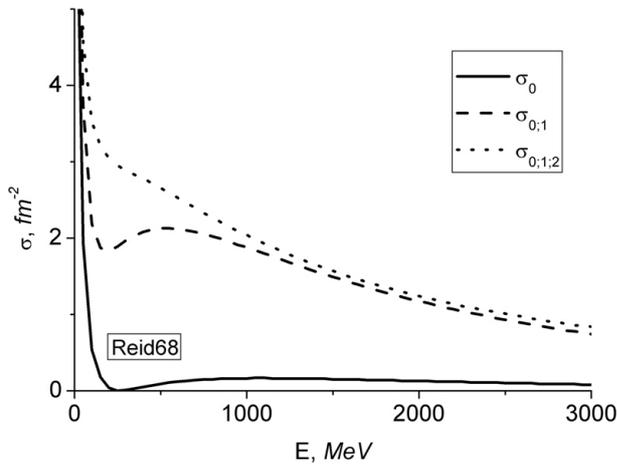

***Fig. 4.*** *Full cross-section of np- scattering for Reid68 potential*

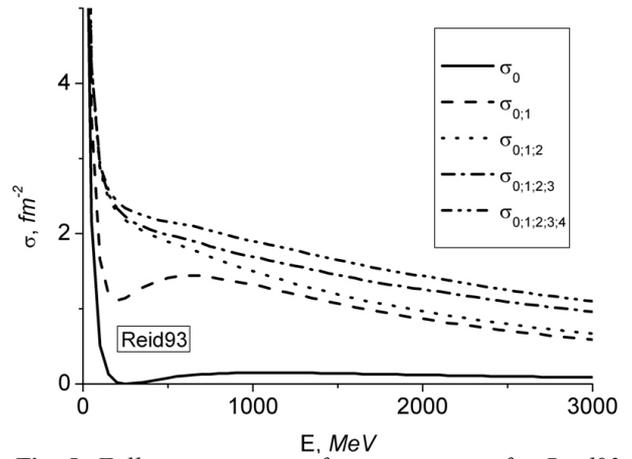

***Fig. 5.*** *Full cross-section of np- scattering for Reid93 potential*

In Fig. 4 and 5 the value $\sigma_l$ is the total cross-section, which calculated through the phase shifts with orbital moment's *l*.

So, results of paper are:

1. The phase-functions method has been used for the first time to calculate *np* phase shifts for the relevant spin configurations in the interval of energies from 1 MeV to 3 GeV for nucleon-nucleon Reid68 and Reid93 potentials.

2. Numerically obtained phase shifts well agree with the Available results of other papers [6, 7] for the same potentials (the deviation makes no more than five percent).

3. The full cross-section has been calculated using the obtained phase shifts on PFM.

This work performed under the grant of the Ministry of Education and Science of Ukraine on the theme of research work of state registration number 0115U001098.

УДК 539.12.01

# РОЗРАХУНОК ФАЗ *NP*- РОЗСІЯННЯ ДО T$_{lab}$=3 ГеВ ДЛЯ ПОТЕНЦІАЛІВ Reid68 І Reid93 ЗА МЕТОДОМ ФАЗОВИХ ФУНКЦІЙ

## В. І. Жаба

*Ужгородський національний університет, 88000, Ужгород, вул. Волошина, 54*

Для обрахунку фаз одноканального нуклон-нуклонного розсіяння розглянуто відомий метод фазових функцій. За допомогою методу фазових функцій чисельно отримано фазові зсуви *np*- розсіяння для $^1S_0$-, $^1P_1$-, $^3P_0$-, $^3P_1$-, $^1D_2$-, $^3D_2$-, $^1F_3$-, $^3F_3$-, $^1G_4$-, $^3G_4$- станів. Розрахунки проведено для реалістичних нуклон-нуклонних потенціалів Reid68 і Reid93. Чисельно розраховані фазові зсуви для енергій до 350 MeB добре узгоджуються з результатами, отриманими іншими методами. По розрахованим фазовим зсувам обчислено повний переріз *np*- розсіяння.

## 1. ВСТУП

Із експериментально спостережуваних величин перерізу розсіяння та енергій переходів отримують у першу чергу інформацію про фази та амплітуди розсіяння, ніж про хвильові функції, що є основним об'єктом дослідження при стандартному підході. Іншими словами, в експерименті спостерігаються не самі хвильові функції, а їх зміни, викликані у результаті взаємодії [1,2]. Тому представляє інтерес отримати і використовувати рівняння, що безпосередньо пов'язують фази й амплітуди розсіяння з потенціалом, не знаходячи при цьому хвильові функції.

Точний розв'язок задачі розсіяння із метою обчислення фаз розсіяння можливий тільки для окремих феноменологічних потенціалів. Коли використовуються реалістичні потенціали, то фази розсіяння обчислюються наближено. Це пов'язано з використанням фізичних апроксимацій або з чисельним розрахунком.

В останні 10 років зріс інтерес до нуклон-нуклонного розсіяння у рамках кіральної теорії збурень [3,4], в послідовному теоретико-польовому підході [5], для парціального хвильового аналізу нижче порогу утворення піона [6]. Також отримуються фази розсіяння через суперсиметрії і факторизації [7], N/D метод обчислення парціальних хвиль для еластичного *NN*- розсіяння [8] або перенормування *NN*- взаємодії для кірального потенціалу двопіонного обміну [9].

До методів розв'язування рівняння Шредінгера з метою отримання фаз розсіяння належать: метод послідовних наближень, борнівське наближення, метод фазових функцій (МФФ) та інші. МФФ виявився досить зручним при розв'язуванні багатьох конкретних задач атомної і ядерної фізики.

Основною і головною перевагою МФФ при застосуванні до задач нуклон-нуклонного розсіяння є та, що МФФ дозволяє отримати фази розсіяння, не знаходячи при цьому хвильові функції як розв'язки рівняння Шредінгера. Завдяки фазовому рівнянню наявний безпосередній зв'язок між фазою розсіяння і потенціалом взаємодії.

Дана робота присвячена розрахунку фазових зсувів *np*- розсіяння у різних спінових станах для реалістичних феноменологічних нуклон-нуклонних потенціалів Reid68 [10] і Reid93 [11] за допомогою МФФ.

## 2. МЕТОД ФАЗОВИХ ФУНКЦІЙ

МФФ - це особливий спосіб розв'язування радіального рівняння Шредінгера

$$u''_l(r) + \left(k^2 - \frac{l(l+1)}{r^2} - U(r)\right)u_l(r) = 0, \quad (1)$$

яке є лінійним диференціальним рівнянням другого порядку. В формулі (1) величина $U(r) = \frac{2m}{\hbar^2}V(r)$ - це перенормований потенціал взаємодії, $m$ - приведена маса. МФФ досить зручний для отримання фаз розсіяння, оскільки по цьому методу не потрібно спочатку обчислювати в широкій області радіальні хвильові функції задачі розсіяння і потім по їх асимптотикам знаходити ці фази.

Стандартний спосіб обчислення фаз розсіяння - це розв'язок рівняння Шредінгера (1) з асимптотичною граничною умовою. МФФ - це перехід від рівняння Шредінгера до рівняння для фазової функції. Для цього роблять заміну [1,2]:

$$u_l(r) = A_l(r)[\cos\delta_l(r) \cdot j_l(kr) - \sin\delta_l(r) \cdot n_l(kr)]. \quad (2)$$

Введені дві нові функції $\delta_l(r)$ і $A_l(r)$ мають зміст відповідних фаз розсіяння і констант нормування (амплітуд) хвильових функцій для розсіяння на визначеній послідовності обрізаних потенціалів. $\delta_l(r)$ і $A_l(r)$ називаються відповідно їх фізичному змісту фазовою й амплітудною функцією. Термін "фазова функція" вперше був використаний у роботі Морзе і Алліса [12]. Рівнянням для фазової функції з початковою умовою є:

$$\delta'_l(r) = -\frac{1}{k}U(r)[\cos\delta_l(r) \cdot j_l(kr) - \sin\delta_l(r) \cdot n_l(kr)]^2, \quad (3)$$

$$\delta_l(0) = 0.$$

Фазове рівняння було вперше отримано Друкарєвим, а потім незалежно у роботах Бергмана,

Колоджеро і Зімека. Частинний випадок рівняння (3) при *l=0* був використаний Морзе і Аллісом при дослідженні задачі *S*- розсіяння повільних електронів на атомах [12].

## 3. РОЗРАХУНКИ ФАЗОВИХ ЗСУВІВ І ОБГОВОРЕННЯ РЕЗУЛЬТАТІВ

Методом фазових функцій чисельно розраховано фазові зсуви *np*- розсіяння для $^1S_0$-, $^1P_1$-, $^3P_0$-, $^3P_1$-, $^1D_2$-, $^3D_2$-, $^1F_3$-, $^3F_3$-, $^1G_4$-, $^3G_4$- станів. Маси нуклонів вибрано такими: $M_p$=938.27231 МеВ; $M_n$=939.56563 МеВ. Був вибраний чисельний метод розв'язання фазового рівняння (3) - метод Рунге-Кутта четвертого порядку [13]. Програмний код для чисельних розрахунків написаний на мові програмування ФОРТРАН. При оптимізованому виборі кроку чисельних розрахунків фазові зсуви отримувалися з точністю до 0.01. Фазові зсуви знаходились при виході фазової функції $\delta_l(r)$ на асимптотику при *r*>25 Фм. Значення фазових зсувів приведено на Рис. 1-3. Фазові зсуви вказано у градусах. Чисельні розрахунки проведено для потенціалів Reid68 і Reid93. Діапазон енергій 1-3000 МеВ.

Для області енергій 1-350 МеВ наявне добре узгодження між фазовими зсувами отриманими на основі МФФ (результати даної роботи) і даними в інших роботах [10,11]). Розходження між результатами становить не більше двох відсотків.

Слід зауважити, що в останні 15 років зріс інтерес для знаходження фазових зсувів при великих енергіях. Нажаль, у літературних джерелах значно менше наявних розрахованих фазових зсувів при великих енергій для потенціалів Рейда. Натомість для інших потенціальних моделей наявні фазові зсуви у широкій енергетичній області. Так фазові зсуви розраховано до 1000 МеВ для нуклон-нуклонних моделей Av18, CD-Bonn і N3LO [14], а також для Arndt, OBEP, Bonn-B, Nijm-3 і Paris [15]. В [16] приведено результати розрахунків фазових зсувів до 1.6 ГеВ для потенціалів інверсії, що базуються на SM94, OSBEP, Av18 і Bonn-B. Крім цього, згідно [17] потенціали Nijm-1, Nijm-2, Av18 та квантової інверсії Джел'фанд-Левітан-Марченко були розширені як *NN* оптичні моделі. Причому був врахований аналіз фазових зсувів по Арндту та ін. (SP00, FA00, WI00) від 300 МеВ до 3 ГеВ. В [18] наявні фазові зсуви тільки до 1000 МеВ для $^1F_3$-стану для потенціалів Nijm-1, Nijm-2, Reid93, Bonn і Av18, які були екстрапольовані для високих енергій.

Крім результатів у згаданих роботах, наявні фазові зсуви отримано до 3 ГеВ для релятивістської оптичної моделі на базі Московського потенціалу [19] та потенціалів MYQ2, MYQ3, MY2, SP07 і Graz II [20], а також до 1.2-3.0 ГеВ для потенціалу Дірака [21] і до 1.2 ГеВ для потенціалу інверсії і Парижського потенціалу [22].

Відмінність між отриманими фазовими зсувами $^1F_3$- стану по МФФ і даними роботи [18] для потенціалу Reid93 становить не більше 5 відсотків. Зрештою розраховані фазові зсуви для конкретної спінової конфігурації при великих енергіях (більше 350 МеВ) для потенціалів Reid68 і Reid93 відрізняються між собою у більшості випадків.

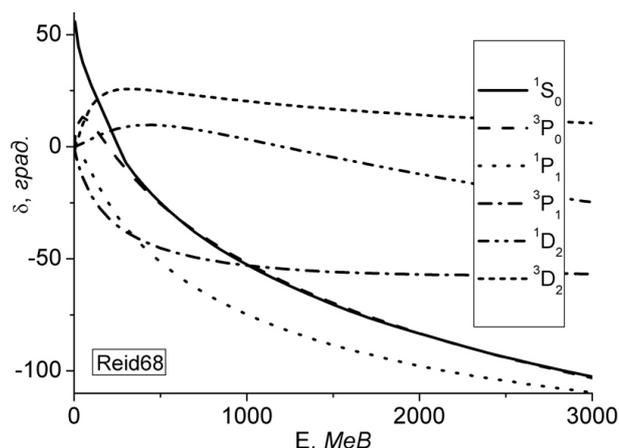

*Рис. 1. Фазові зсуви np- розсіяння для потенціалу Reid68*

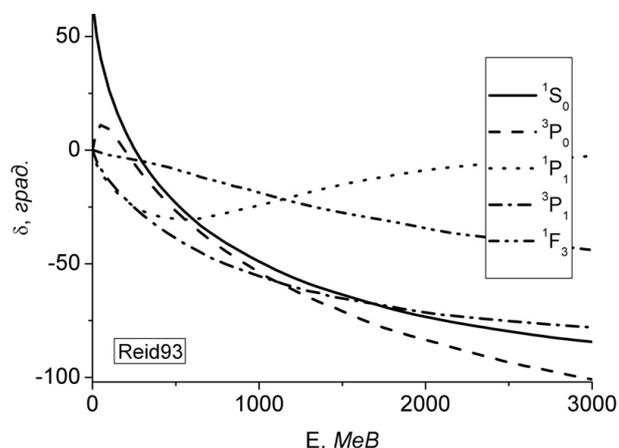

*Рис. 2. Фазові зсуви np- розсіяння для потенціалу Reid93*

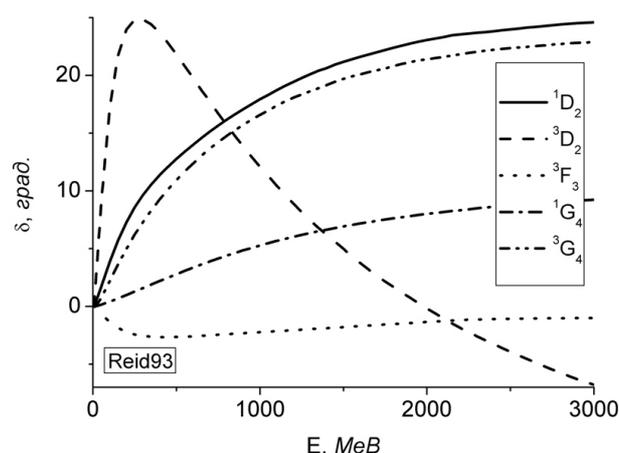

*Рис. 3. Фазові зсуви np- розсіяння для потенціалу Reid93*

Якщо ж порівнювати отримані фазові зсуви по МФФ при великих енергіях для потенціалів Рейда з даними для вказаних інших потенціальних моделей, то очевидне розходження між ними. Звісно, що це пов'язано з особливостями структури нуклон-нуклонних потенціалів.

Незважаючи на очевидні поліпшення в описі даних при енергіях нижче 400 МеВ, теоретичні розрахунки досі показують деякі значні систематичні недоліки [22]. Зокрема, повністю залишається незрозумілою розбіжність спінових спостережуваних при передачах імпульсу нижче 1 Фм$^{-1}$. Походження таких розбіжностей з даними можна було б віднести до феноменологічного обмеження «голого» нуклон-нуклонного потенціалу особливо при більш високих енергіях, а також і до спрощень в моделі для NN ефективної взаємодії або до того факту, що оптична модель потенціалу тільки була розроблена до найнижчого порядку. Вище 350-400 МеВ розбіжності між потенціалами NN взаємодії стають ще більш очевидними. При енергіях більше 400 МеВ NN потенціали мають прикладне застосування. Тому оцінка теорії потребує більш точного опису «голої» взаємодії двох нуклонів.

Поряд з фазовими зсувами у задачах розсіяння приходиться мати справу з амплітудами розсіяння, елементами S- матриці і цілим рядом інших параметрів. По відомим фазам розсіяння отримують повну амплітуду, повний переріз і парціальну амплітуду розсіяння відповідно [1]

$$F(\theta) = \frac{1}{k}\sum_{l=0}^{\infty}(2l+1)e^{i\delta_l}\sin\delta_l P_l(\cos\theta), \quad (4)$$

$$\sigma = \frac{4\pi}{k^2}\sum_{l=0}^{\infty}(2l+1)\sin^2\delta_l, \quad (5)$$

$$f_l = \frac{1}{k}e^{i\delta_l}\sin\delta_l, \quad (6)$$

де $P_l(\cos\theta)$ - поліноми Лежандра, $\theta$ - полярний кут.

У роботі [23] вказаний повний переріз розсіяння, розрахований по фазових зсувах при енергіях 1-350 МеВ для потенціалів Неймегенської групи (зокрема і для Reid93) і для потенціалу Av18. Використовуючи фазові зсуви при енергіях до 350 МеВ, результати розрахунків параметрів Вольфенштейна як функції кута ц.м. та E$_{LAB}$ наведені в [24] для PWA і Av18, а в [25] - для Reid93 та NijmII.

Результати розрахунків повного перерізу np- розсіяння (5) приведено на Рис. 4 і 5.

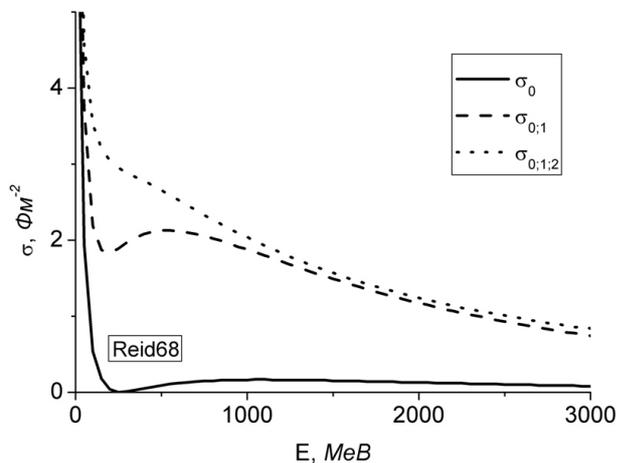

**Рис. 4.** Повний переріз np- розсіяння для потенціалу Reid68

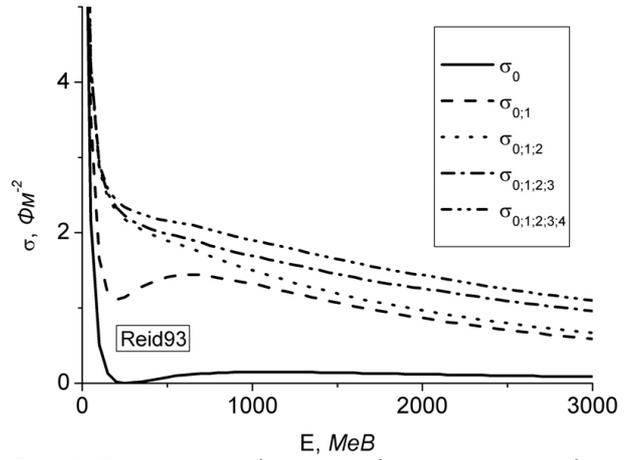

**Рис. 5.** Повний переріз np- розсіяння для потенціалу Reid93

На Рис. 4 і 5 величина $\sigma_l$ - це повний переріз, розрахований по фазовим зсувам з орбітальним моментам l.

Отже, результатами роботи є:

1. Вперше методом фазових функцій розраховано np фазові зсуви для відповідних спінових конфігурацій в області енергій від 1 МеВ до 3 ГеВ для нуклон-нуклонних потенціалів Reid68 і Reid93.

2. Чисельно отримані фазові зсуви добре узгоджуються з наявними результатами інших робіт для цих же потенціалів (відхилення становить не більше 5 відсотків).

3. По розрахованим фазовим зсувам за МФФ обчислено повний переріз np- розсіяння.

## Література